\documentclass[showpacs,twocolumn,prl]{revtex4}

\usepackage{amsmath}
\usepackage{amssymb}
\usepackage[T1]{fontenc}
\usepackage{graphicx}

\newcommand{\barf}{\bar{f}}
\newcommand{\kb}{k_{{\scriptscriptstyle B}}}

\newcommand{\pt}{\, .}
\newcommand{\vir}{\, ,}
\newcommand{\as}{a_\sigma}
\newcommand{\El}{{\scriptscriptstyle El}}
\newcommand{\HC}{{\scriptscriptstyle{HC}}}
\newcommand{\nDH}{{\scriptscriptstyle{\rm{DH}}}}
\newcommand{\nMC}{{\scriptscriptstyle{\rm{MC}}}}

\newcommand{\Ex}{{\scriptscriptstyle{\rm{Ex}}}}
\newcommand{\nE}{{\rm{E}}}

\newcommand{\Tce}{T_c^*}
\newcommand{\roce}{\rho_c^*}

\newcommand{\pe}{\! = \!}

\newcommand{\capms}{C$_z$APMs }
\newcommand{\capm}{C$_z$APM }
\newcommand{\bI}{\textbf{I} }
\newcommand{\fbel}{\bar{f}^{{\scriptscriptstyle{\rm{El}}}}}
\newcommand{\fbhc}{\bar{f}^{{\scriptscriptstyle{\rm{HC}}}}}
\newcommand{\cE}{\mathcal{E}}
\newcommand{\vecr}{\mathbf{r}}
\newcommand{\mz}{{m,z}}
\newcommand{\vare}{\varepsilon}
\newcommand{\roh}{\hat{\rho}}

\begin{document}

\title{How Multivalency controls Ionic Criticality}

\date{\today}

\author{Michael E. Fisher}

\author{Jean-No\"el Aqua}

\altaffiliation{Present address: \'Ecole normale supérieure de Lyon, 69364,
Lyon, France}

\author{Shubho Banerjee}

\altaffiliation{Present address: Department of Physics, Rhodes College, Memphis
TN 38112}

\affiliation{Institute for Physical Science and Technology, University of
Maryland, College Park, Maryland 20742, USA}

\begin{abstract}

To understand how multivalency influences the reduced critical temperatures,
$\Tce (z)$, and densities, $\roce (z)$, of $z\! : \! 1$ ionic fluids, we study
equisized hard-sphere models with $z \pe 1\!-\!3$. Following Debye, Hückel
and Bjerrum, association into ion clusters is treated with, also, ionic
solvation and excluded volume. In good accord with simulations but
contradicting integral-equation and field theories, $\Tce$ \emph{falls}
when $z$ increases while $\roce$ \emph{rises} steeply: that $80-90\%$
of the ions are bound in clusters near $T_c$ serves to explain these
trends. For $z \! \neq \! 1$ interphase Galvani potentials arise and are
evaluated.

\end{abstract}


\maketitle

Multivalent ions play a significant role in condensed-matter, physicochemical, biophysical and,
via the plasma transition, astrophysical contexts \cite{multi}.  The effects of multivalency are,
however, often hard to comprehend. One central issue---relevant to electrolyte solutions, molten
salts, liquid metals, and dense plasmas \cite{multi}---arises in Coulomb-driven
phase separation. The most basic model for
such ionic fluids consists of $N \pe \rho
V$ hard-core spherical ions of various species $\sigma$ in a volume  $V$  of
uniform dielectric constant $D$,  with $N_\sigma \pe \rho_\sigma V$  ions of
diameter $a_\sigma$   carrying charges $q_\sigma \pe z_\sigma q_0$, where $q_0$
is an elementary charge.  In the simple equisized  $z$:$1$
\emph{charge asymmetric primitive models} (C$_z$APMs), on which we focus here, one
has $\sigma \pe +, -$, $a_+ \pe a_-$, and $q_+ \pe z q_0$, $q_- \pe - q_0$. The
basic energy scale and associated reduced temperature and density
are then $\varepsilon \pe  z q_0^2 / D a$, $T^* \pe  \kb T / \varepsilon$,
$\rho^* \pe \rho a^3$.
\phantom{\cite{netz&orla99}}

\begin{figure}[t]
\includegraphics[width=7cm,height=6cm]{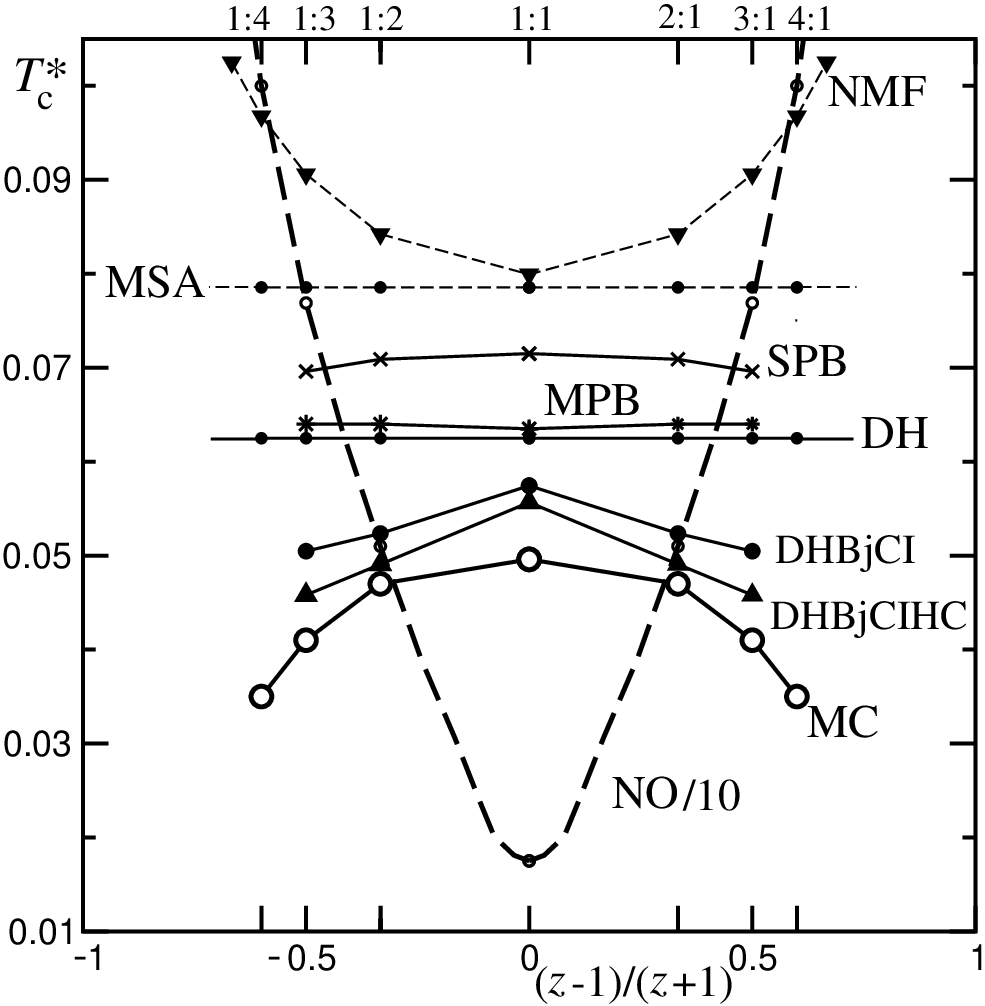}
\caption{\label{Tc2x}
Reduced critical temperatures for $z$:$1$ charge
asymmetric equisized hard-core primitive model electrolytes (\capms) according
to Monte Carlo (MC) simulations \cite{cp99} ; Debye-H\"uckel (DH) theory;
field-theoretic approaches : NO \cite{netz&orla99} (with a factor
$\frac{1}{10}$) and NMF \cite{cail05}; approximate integral equations : MSA
\cite{wais&lebo72}, SPB, and MPB \cite{sabi&bhui98c}; and the present DHBjCI
and DHBjCIHC solvated ion-cluster theory \cite{abf04}; See text.}
\end{figure}

\begin{figure}[t]
\includegraphics[width=7cm,height=6cm]{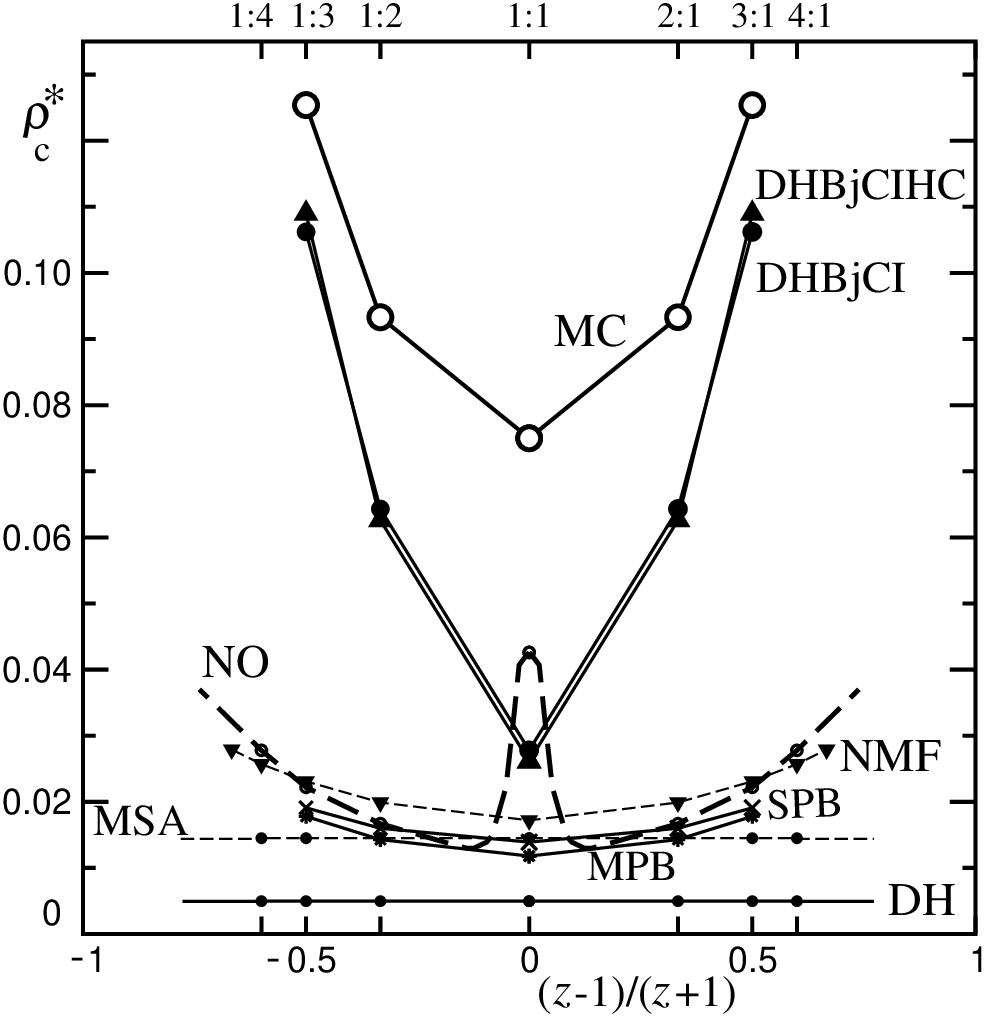}
\caption{\label{roc2x}  Reduced critical densities $\roce (z)$, for the \capm electrolyte as in
Fig.~\ref{Tc2x} (except that the NO plot is not rescaled).}
\end{figure}

Monte Carlo simulations \cite{cp99} show that (at least for $z \!  \lesssim \!
5$)  the \capms exhibit ``gas-liquid'' phase separation; furthermore, the
critical parameters, $\Tce (z)$ and $\roce (z)$, are found to reasonable
precision : see Table \ref{tablei} and the open circles in Figs.~\ref{Tc2x} and
\ref{roc2x}. One observes that $\Tce (z)$ \emph{falls}
with increasing $z$,  while $\roce (z)$
\emph{rises} sharply. But we
ask : How can these trends be understood? Or accounted for
semiquantitatively?  To address this issue we review briefly
previous work, including a pioneering field-theoretic attack
\cite{netz&orla99}, and then report on a recent study \cite{abf04} which we
believe provides significant insight. This extends an earlier analysis \bI
\cite{fl} for the symmetric  $z \pe 1$ \emph{restricted primitive model} (RPM)
that was founded on the original Debye-Hückel (DH) approach but incorporated (i)
Bjerrum ion pairs and (ii) their \emph{solvation} in the residual ionic fluid.
For $z \pe 2$ and $3$ larger ion clusters, trimers and tetramers, must be
included \cite{abf04,arty&kobe03c}; but then explicit results are also obtained
for the \emph{interphase Galvani potential} \cite{spar72l} that appears in any
two-phase nonsymmetric ionic system \cite{abf04,spar72l}.

\begin{table}[h]
\caption{\label{tablei} Monte Carlo (MC) estimates \cite{cp99} for the reduced critical parameters for
$z$:$1$ equisized hard-sphere electrolytes, values calculated from DHBjCIHC theory (CI) \cite{abf04},
and approximate estimates based on ion cluster statistics : see text.}
\begin{ruledtabular}
\begin{tabular}{llllllllll}
 & \multicolumn{2}{l}{critical temp.} & \multicolumn{2}{l}{$10^2 \Tce (z)$} & &
  \multicolumn{2}{l}{critical density} & \multicolumn{2}{l}{$10^2 \roce (z)$} \\
 $z$ & MC & CI & $\nE_\nDH$ & $\nE_{{\nMC}}$ & &
           MC & CI & $\nE_\rho$ & $\nE_\kappa$ \\
  $1$      & $4.93_3$ & $5.56_9$ & $5.45$ & $4.93_5$ & & $\phantom{1}7.50$ & $2.61_4$ & $2.72$ & $2.37$ \\
  $2$      & $4.70$ & $4.90_7$ & $5.11$ & $4.65$ & & $\phantom{1}9.3$ & $6.26_1$ & $4.27$ & $3.49$ \\
  $3$      & $4.10$ & $4.33_4$ & $4.85$ & $4.44$ & & $12.5$ & $11.90$ & $6.96$ & $5.40$ \\
\end{tabular}
\end{ruledtabular}
\end{table}

The field-theoretic analysis of Netz and Orland (NO) \cite{netz&orla99} was designed to
address  $z$:$1$ ionic fluids and colloids ($z \gg 1$)
and to include correlations in a systematic manner.  The Coulomb interaction,
$q_\sigma q_\tau /r$,
was transformed to yield a functional integral over an auxiliary potential
$\phi(\mathbf{r})$. At the $\langle \phi^2 \rangle$ level the DH effective
interaction, $v_{{\nDH}} \varpropto e^{-\kappa r}/r$ is captured with
\begin{equation} \label{kappa}
  \kappa^2 (T; \{\rho_\sigma \}) = 4 \pi (q_0^2 / D \kb T) \sum\nolimits_\sigma
  z_\sigma^2 \rho_\sigma \pt
\end{equation}
The reduced free energy density,
$\bar{f} (T; \rho) \! \equiv \! - F / V \kb T$, was computed to eighth order in
$\phi$ but a momentum cut-off is essential:
NO adopted $|\mathbf{k}_\Lambda| \pe 2 \pi /a$
thereby incorporating the ionic diameter and, for the $z$:$1$ case, leading to
$\kappa^2 a^2 \pe 4 \pi \rho^* / T^*$. Since this treatment of the hard
cores is approximate, accurate predictions for $\Tce (z)$  and
$\roce (z)$ are not expected.  Nevertheless, one might
anticipate reliable  \emph{trends} when $z$ varies in contrast to DH theory
which yields \emph{no dependence} on $z$  with (after \textbf{I})
\begin{equation}
        \label{DH}
        {\rm{DH}} : \quad \kappa_c a = 1, \quad \Tce = \mbox{$\frac{1}{16}$},
        \quad \roce = 1/64 \pi \simeq 0.005 \pt
\end{equation}
In fact, as NO report, ``the [predicted] deviations from DH theory are
pronounced'' for $z \! > \! 1$ :  see the bold dashed plots in Figs.~\ref{Tc2x}
and \ref{roc2x}.

But evidently the NO results are not merely quantitatively
wrong; the \emph{trends}
are quite incorrect since $\Tce$  is asserted to rise rapidly (instead of
falling) while $\roce$ falls sharply for small $z \!- \! 1$  (instead of rising)
and \emph{then}  increases but much too slowly. While one may blame
the approximate treatment of the hard cores, we believe this
is \emph{not} the primary culprit.  Indeed, a recent field-theoretic analysis
paid closer attention to the ion-ion repulsions \cite{cail05}; but the
subsequent ``new mean-field'' (NMF) results still
exhibit strong increases in $\Tce$ and an overly weak variation of $\roce$:  see
the NMF plots in Figs.~\ref{Tc2x} and \ref{roc2x} \cite{cail05}.

Integral equation theories are hardly better : see Figs.~\ref{Tc2x} and
\ref{roc2x}. The \emph{mean spherical approximation}
(MSA), like DH theory, predicts \emph{no variation} of
$\Tce$ and $\roce$  with $z$ \cite{wais&lebo72}. A \emph{symmetric
Poisson-Boltzmann} (SPB) theory \cite{sabi&bhui98c} \emph{does} predict the
correct falling and rising trends for $\Tce$ and $\roce$,  but the degree of
variation is woefully inadequate. Moreover, the
\emph{modified Poisson-Boltzmann} (MPB) approximation, that the same authors
\cite{sabi&bhui98c} argue should be more reliable, yields the
\emph{wrong} trend for $\Tce$.

In order to better understand the effects of multivalency we turn to recent calculations
\cite{abf04,arty&kobe03c} based on the \emph{solvated ion-cluster}
view \cite{fl} of the \capm near criticality that is supported `pictorially' by
simulations \cite{cp99}. In brief, the aim is to construct the free energy
density, $\bar{f} (T; \{\rho_\sigma\})$, for ionic species $\sigma$ consisting
(i) of  $+$  and  $-$ \emph{monomers}, i.e., isolated, $n_+ \pe n_- \pe 1$
single, unassociated ions of valency $z_+ \pe z$ and $z_- \pe -1$ ; (ii)
a set of associated \emph{primary} clusters, $\sigma \pe 2$, $3$, $\ldots$,
dimers, trimers, $\ldots$, each consisting of one ``central'' $+$ ion and
$m_\sigma \pe \sigma -1$ ``satellite'' counterions for a total of $n_\sigma \pe
m_\sigma + 1$ ions in a cluster of valency $z_\sigma \pe z - m_\sigma$; up to
(iii), the largest primary cluster, the \emph{neutral} or
`molecular'  $(z+1)$-mer of one $z_+$ ion and $z$ negative ions
\cite{abf04,arty&kobe03c}.

For each species, $\bar{f}$ contains an ideal-gas
term $\bar{f}^{{\scriptscriptstyle{\rm Id}}} (T, \rho_\sigma)$, and an
electrostatic term $\fbel_\sigma (T, \{\rho_\tau\})$, that,
following DH, incorporates \emph{{\bf C}luster solvation} in the partially
associated \emph{{\bf I}onic fluid}: this description is thus dubbed
``DHBjCI'' \cite{abf04}. By adding a \emph{{\bf H}ard {\bf C}ore}
free-volume term, $\fbhc (\{ \rho_\sigma\})$, as in \textbf{I}, one
may also account for those excluded
volume effects not \emph{already} encompassed in the basic
solvation and association calculations \cite{abf04,fl},
so generating a
``DHBjCIHC'' theory \cite{abf04}. (The effective HC virial coefficient
$B_\sigma^{{\rm{bcc}}} \pe$
$4 {a_{\sigma}}^{\! \! \! \!3} / 3^{3/2}$ has been adopted \cite{abf04,fl}.)
Examination of Figs.~\ref{Tc2x} and \ref{roc2x} reveals that these
solvated ion-cluster theories
are surprisingly successful! Not only are \emph{both} the downward
trend in $\Tce (z)$ \emph{and} the rapid rise of $\roce (z)$ well
captured, but the \emph{quantitative} agreement with each of the
MC estimates is significantly better than achieved by other approaches.

\begin{figure}[t]
\includegraphics[width=7cm,height=6cm]{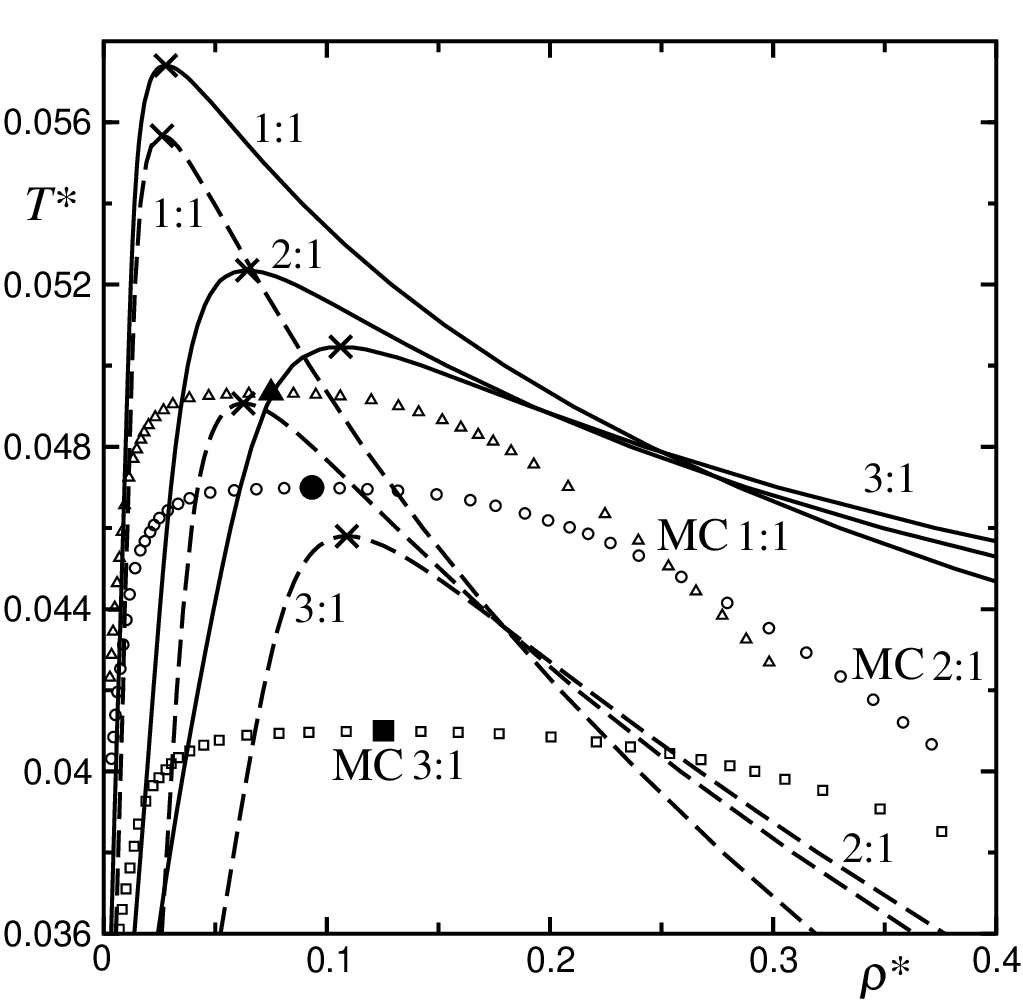}
\caption{\label{diagphas} Coexistence curves predicted for $z$:$1$ equisized
primitive models by the DHBjCI and DHBjCIHC theories (solid and dashed lines,
respectively) together with Monte Carlo estimates (MC) based on
\cite{young&mef&luij03}.}
\end{figure}

One must recognize that (all) these theories are
of mean-field character: thus $5$ to $15\%$ over-estimates of $\Tce (z)$
are to be expected. Indeed, neglected fluctuations typically depress
$T_c$ by such amounts and also flatten the \emph{coexistence
curves} as seen in Fig.~\ref{diagphas}. Second, note that the hard-core terms
have a small effect on $\roce (z)$
while reducing $\Tce (z)$ values by only $5\!-\!10\%$. Nevertheless,
Fig.~\ref{diagphas} reveals that the liquid phases, especially for
$\rho^* \gtrsim 0.15$, are sensitive to
$\fbhc$: but, recall the discussion in \textbf{I}. In fact, the crucial feature
of DHBj-type theories---not represented in field-theoretic or standard
integral-equation treatments---is the \textit{chemical equilibrium} maintained
between the cluster species via the Law of Mass Action:
\begin{equation}
\label{eqchi}
 \rho_\sigma = K_{m,z} (T) \, \rho_+ \, \rho_-^{m} \, \exp \left[ \mu^{\Ex}_+ +
        m \mu^{\Ex}_- - \mu_\sigma^{\Ex} \right] \vir
\end{equation}
for $\sigma \pe m + 1 \geq 2$, with the excess chemical potentials
$\mu_\sigma^{\Ex} = - (\partial / \partial \rho_\sigma) [\barf^{\HC} +
                \sum_\sigma \barf_\sigma^{\El} ]$,
while \textit{the association constants} are taken as \cite{abf04,fl}
\begin{equation}
        \label{Kmz}
        K_{m, z} (T; R) = \frac{1}{m!} \prod_{i=1}^m \int_a^R d \vecr_i \exp \left(
                - \frac{\cE_\mz (\{\vecr_i \})}{\kb T}  \right) \vir
 \end{equation}
in which $\cE_\mz (\{\vecr_i\})$ is the electrostatic energy of an isolated
$(m+1)$-mer with satellite coordinates $\{\mathbf{r}_i\}$. The lower limits
$a$ and the condition $\cE_{m,z} \pe + \infty$ for $|\vecr_i - \vecr_j| < a$,
represent hard cores. Following Bjerrum \cite{fl}, the necessary cut-off radius $R$
is chosen so that $(\partial K_\mz / \partial R)$ is minimal. The resulting
$3$-fold $K_{2,z}$ integral is managable but the 6-fold integral
for $K_{3,3}$ requires a Pad\'e approximant study of the low-$T$ expansion
cross-checked to a part in $10^3$ by MC evaluations \cite{abf04}. It transpires,
however, that $\Tce$ and $\roce$ are \textit{insensitive}
to the $K_\mz$ values \cite{abf04}.

Lastly, one needs to account for the solvation of \emph{all} the ion species,
$\sigma$, by the free ions and charged clusters via the electrostatic terms
\cite{abf04,fl,arty&kobe03c}
\begin{equation}
  \label{fels}
        \barf_\sigma^{\El} (T; \{ \rho_\tau\}) = \frac{4 \pi \rho_\sigma}{D \kb T} \sum_{l=0}^\infty
        \frac{u_{2l} (\kappa \as) }{\as^{2l+1}} \sum_{m=-l}^{l}
         \langle |Q_{l m}^\sigma|^2  \rangle \vir
\end{equation}
where the $u_{2l} (x)$ are
related to the spherical Bessel functions $k_l (x)$ \cite{abf04}; the second sum
requires the cluster electric multipole moments, $Q_{l,m}^\sigma$, thermally
averaged \cite{abf04} over the ionic configurations that already enter
in the $K_\mz (T)$.

Finally, $a_\sigma$ is an \textit{effective cluster diameter}, i.e. the
radius of the approximating sphere (centered to minimize $\barf_\sigma^\El$)
that substitutes for the true, thermally fluctuating, hard-core exclusion
domain: see \bI and \cite{abf04}. One concludes, as in \textbf{I}, that a most
reasonable choice for $a_\sigma$ is the average over solid angle of the radial
distance to the true exclusion surface of the ground-state cluster : this yields
$a_2 \pe (\frac{3}{4} + \frac{3}{8} \ln 3 \simeq 1.162 ) a$, $a_3 \pe 1.250 a$
and $a_4 \pe 1.375 a$. For $z \pe 1$ the values of $\Tce$ and $\roce$ vary by
less than $\pm 2\%$ over plausible alternatives for  $a_2$ \cite{abf04}; but the
sensitivity to $a_3$ and $a_4$ for $z \pe 2$ and $3$ is greater. As a result,
this hard-to-avoid approximation contributes significantly to the overall
quantitative uncertainties.

From the total free energy $\barf (T, \{ \rho_\sigma\})$, all thermodynamic
properties follow \cite{abf04,fl}.
One may then conclude from Figs.~\ref{Tc2x} and \ref{roc2x}
that the principal defect of the field-theoretic and integral-equation
approaches is a failure to account effectively for strong ionic association near
criticality. But can the actual \textit{trends} of $\Tce$ and
$\roce$ with $z$ be demonstrated in a direct, transparent way? To answer,
consider the \textit{fractions}, $y_\sigma \pe n_\sigma N_\sigma / N$, of ions
bound in clusters of $n_\sigma$ ions with $\rho_\sigma \pe
(y_\sigma / n_\sigma) \rho$. The critical point values  that
result from DHBjCIHC theory \cite{abf04,foot} are displayed in Table
\ref{tableii}. A significant fact is the rapid decrease in $y_+^c$, the
fraction of unassociated $z_+$ ions, from $9.1$ to $1.3$ to $0.3\%$.
But more can be learned!

\begin{table}[t]
\caption{\label{tableii} Inverse screening length $\kappa$ and fractions,
$y_\sigma \pe n_\sigma N_\sigma / N$, of ions in clusters of $n_\sigma$ ions at criticality, as percentages,
according to DHBjCIHC theory \cite{abf04}. }
\begin{ruledtabular}
\begin{tabular}{lllllll}
 $z$ & $\kappa_c a$ & $y_+^c$ & $y_-^c$ & $y_2^c$ & $y_3^c$ & $y_4^c$ \\
 1     & $1.04$           & $9.14$    & $\phantom{1}9.14$                & $81.72$         & $-$        & $-$ \\
 2     & $1.37$           & $1.31$    & $10.33$                                 & $15.43$         & $72.93$        & $-$ \\
 3     & $1.57$           & $0.34$    & $\phantom{1}8.04$   & $\phantom{1}3.32$     & $11.13$        & $77.17$ \\
\end{tabular}
\end{ruledtabular}
\end{table}

To understand  the variation of $\Tce (z)$ let us regard the electrolyte in the
critical region as a mixture of clusters with fixed mole fractions
$x_\sigma  \pe (y_\sigma / n_\sigma) / \sum_\tau (y_\tau/n_\tau)$. A pair
$(\sigma, \tau)$ will either mutually repel
\textit{or} attract with pairwise binding energies, say, $\vare_{\sigma
\tau}$. Thus unlike monomers attract with $\vare_\pm \pe \vare$. However, a
\textit{dimer} attracts only \textit{negative monomers} with $\vare_{2 -} \pe (z
- \frac{1}{2}) \vare / z$; but repels all $z_+ \! \geq \! +2$ ions. Two dimers
repel when $z \! \geq \! 3$; but one has $\vare_{2,2}/\vare \simeq 0.586$ and
$0.345$ for $z \pe 1$ and $2$. And so on.

To estimate $\Tce$  for this mixture we adopt a van-der-Waals
approach as in 
[2(d)]. Thus, for the overall cluster
density $\roh \, \, (= \rho \sum_\sigma y_\sigma / n_\sigma)$, we take $p / \roh
\kb T \simeq Z (B_0 \roh) + B_1(T^*) \roh$ with $Z(u) = 1+u+\ldots$
in which the second virial coefficient has been decomposed as $B(T^*) \pe B_0 + B_1 (T^*)$ where
$B_0$ ($=b_0 a^3$, say) represents the hard-core repulsions while $B_1 (T^*)$ embodies the attractions.
Solving $\partial_\rho p \pe \partial_\rho^2 p \pe 0$, as usual, yields $\roce$
and $B_c^* \! \equiv \! B_1 (\Tce) / b_0 a^3$. At low $T$, which is relevant here, one has
\begin{equation}
        \label{B1}
        B_1 (T^*) \approx - \sum\nolimits_{\sigma, \tau} b_{\sigma \tau} a^3
x_\sigma x_\tau \exp(\vare^*_{\sigma \tau}                 / T^*) \vir
\end{equation}
where $\vare_{\sigma \tau}^* \! \equiv \! \vare_{\sigma \tau} / \vare$, while $b_{\sigma \tau} a^3$
specifies the volume of mutual attractions: this \textit{vanishes} if $\sigma$ and $\tau$ repel.

Now, the $x_+ x_-$ term dominates in $B_1 (T^*)$ at low $T$ with corrections
of relative order $(x_2^2/x_+ x_-) e^{-0.414/T^*}$ for $z \pe 1$ and $2
(x_2/x_+) e^{-1/2zT^*}$ for $z \geq 2$. We may then \textit{calibrate} $B_1
(\Tce) / a^3$ by using pure DH theory \eqref{DH} for which, since association is
not considered, $x_+ \pe x_- = \frac{1}{2}$. Thereby we obtain the $\nE_\nDH$
estimates
\begin{equation}
        \label{edh}
        \Tce (z) \simeq 1/(16+|\ln 4 x_+^c (z) x_-^c (z)|) \vir
\end{equation}
in which $x_+^c  \! \varpropto \! y_+^c$ and $x_-^c \! \varpropto \! y_-^c$
follow from Table \ref{tableii}.

The resulting predictions are listed in Table \ref{tablei} under
${\rm{E}}_\nDH$. In light of the heuristic nature of the
arguments, they reflect the trend of the MC and CI values surprizingly well.
Certainly the contention that association is a prime factor is well confirmed.
By replacing 16 by 20.27 (or 17.96) in \eqref{edh}, and the factor $4$ by
$1/x_+^c (1) x_-^c (1)$, one calibrates $B_1 (\Tce)$ on the MC (or CI) values
for the RPM. Column ${\rm{E}}_{{\scriptscriptstyle{\rm{MC}}}}$ in Table
\ref{tablei} lists the MC-calibrated values : for $z \pe 2$ and $3$ these match
the Monte Carlo estimates to within $1\%$ and $8\%$, respectively \cite{abff2}.

Now, for the critical density, the significance of ion pairing is
already clear in \textit{pure} DHBj theory for the RPM \cite{fl}. The heavy
depletion of the free ions (which, in DHBj theory, drive the transition
\emph{alone}) means that to reach criticality the overall density $\rho \, (= \!
\rho_+ \! + \! \rho_- \! + \! 2 \rho_2)$ must be increased until the DH
criterion $\rho_+^* + \rho_-^* \pe \rho^{* c}_\nDH \pe 1/64\pi$ is met : see
\eqref{DH}. Does the same depletion-by-association mechanism account for the
$z$-dependence of $\roce (z)$?

To progress, rewrite \eqref{kappa} generally as
$\kappa^2 a^2 \pe 4 \pi \rho^\dagger / T^*$, with the
\textit{effective, depleted ionic density}
\begin{equation}
\label{rod}
        \rho^\dagger \equiv \rho^* \sum\nolimits_\sigma z_\sigma^2 y_\sigma (T; \{ \rho_\sigma\})/ z n_\sigma \pt
\end{equation}
If one accepts the DH criterion and uses Table \ref{tableii},
the estimates  $\nE_\rho$, in Table \ref{tablei}, result. Although
these fall short of the Monte Carlo values by $74$, $54$ and $44\%$ for
$z \pe 1\!-\!3$, they reproduce the accelerating increase with $z$ (by
factors $1.57$,  $1.63$ vs. $1.24$, $1.34$).

An alternative approach adopts the DH value $\kappa_c a \pe
1$: see \eqref{DH} but note from Table \ref{tableii} that DHBjCIHC theory
implies that $\kappa_c a$ rises from $1.04$ for the RPM to $1.57$ for $z \pe 3$.
Then using the $\nE_\nDH$ values for $\Tce$, in Table \ref{tablei}, leads to the
$\nE_{\kappa}$
predictions for $\roce (z)$ :  these are all rather low
but the increases with $z$, by factors $1.47$, $1.55$, again reflect the
correct behavior.

\begin{figure}
\includegraphics[width=7cm,height=6cm]{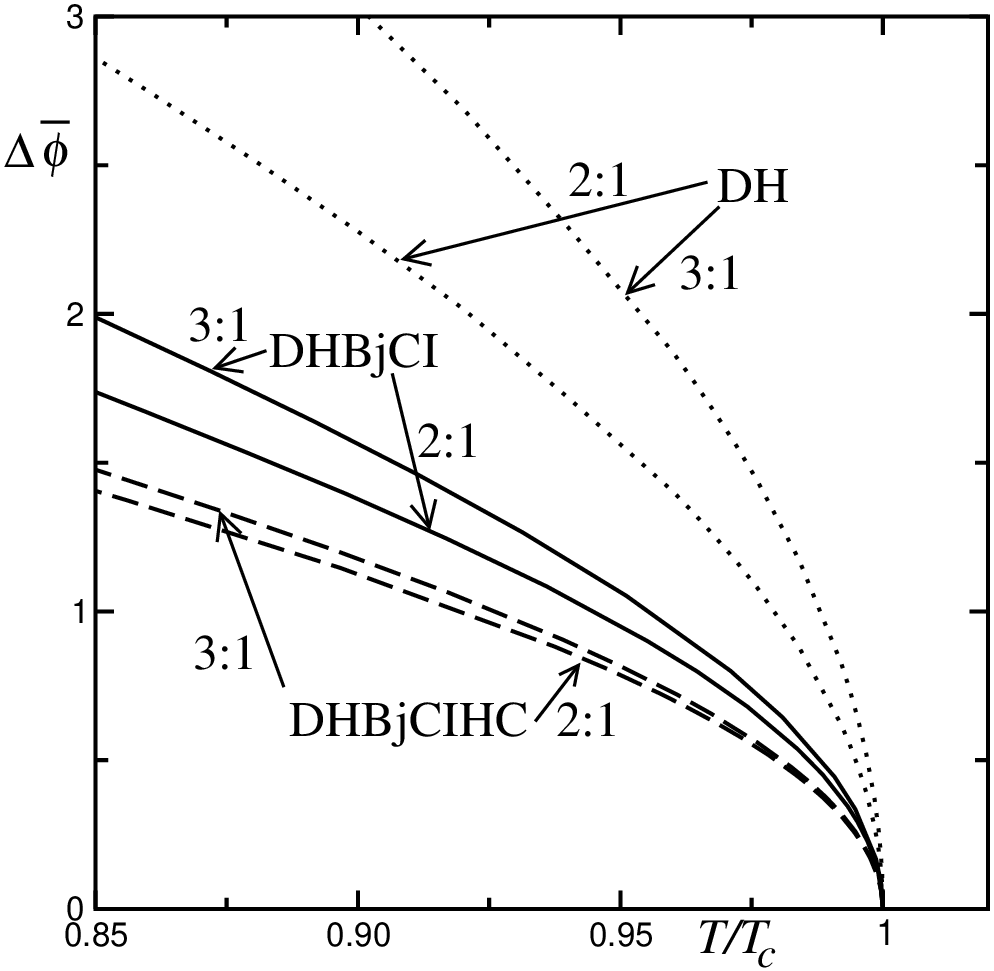}
  \caption{\label{galvani} Reduced Galvani potentials, $\Delta \bar{\phi} \pe q_0 \Delta \phi / \kb T$, vs.
  $T/T_c$ for $z$:$1$ electrolytes according to pure DH theory (dotted) and DHBjCI(HC) theories : solid
  (dashed) plots. }
\end{figure}

Finally, we note that the
Galvani potential, $\Delta \phi$, that arises between coexisting phases in
charge asymmetric fluids is readily calculated \cite{abf04,spar72l}.
The predictions from pure DH theory are shown dotted in
Fig.~\ref{galvani} : one finds $\Delta \phi_\nDH \varpropto (1-z^{-1})$. The
other plots result from the DH\-Bj\-CI and
DHBjCIHC theories \cite{abf04}. Surprizingly, the
calculations suggest no clear trend with $z$. It is natural to conjecture
that $\Delta \phi$ vanishes as $G_0 (T_c - T)^\beta$; moreover to the
extent that the expected mean-field value $\beta \pe \frac{1}{2}$ is realized,
the present results support this.

In conclusion, we have elucidated the mechanisms underlying how multivalency influences critical
behavior. Specifically, we have summarized briefly analytical calculations
for $3$:$1$, $2$:$1$ and $1$:$1$ equisized charged hard-sphere fluids
\cite{abf04} that, for the first time, reasonably reflect the true variation of
critical temperatures and densities, $\Tce (z)$ and $\roce (z)$ (as revealed by
simulations \cite{cp99}). On that basis, supported by analysis
that correlates $\Tce (z)$ and $\roce(z)$ with the increasingly depleted
populations of free ions and charged clusters as $z$ increases, it is clear that
recognizing ionic association is inescapable for a successful theory. Previous
treatments \cite{netz&orla99,cail05,wais&lebo72,sabi&bhui98c}, lacking allowance
for ion clusters fail seriously. The ion-cluster solvation theories also yield
quantitative results for the interphase Galvani potentials.

\begin{acknowledgments}
National Science Foundation support via Grants
CHE 99-81772 and 03-01101 is gratefully acknowledged.
\end{acknowledgments}


\end{document}